# Regional innovation systems in Hungary:

# The failing synergy at the national level

Balázs Lengyel * and Loet Leydesdorff **

\* Institute for World Economics, Hungarian Academy of Science, Országház u. 30, H-Budapest, Hungary; e-mail: blengyel@vki.hu;

\*\* Amsterdam School of Communications Research (ASCoR), University of Amsterdam, Kloveniersburgwal 48, NL-1012 CX Amsterdam, The Netherlands; e-mail: loet@leydesdorff.net; http://www.leydesdorff.net .

**Abstract**

We use entropy statistics in this paper to measure the synergies of knowledge exploration, knowledge exploitation, and organizational control in the Hungarian innovation system. Our data consists of high-tech, medium-tech firms, and knowledge-intensive services categorized in terms of sub-regions (proxy for geography), industrial sectors (proxy for technology), and firm size (proxy for organization). Configurational information among these three dimensions is used as an indicator of the reduction of uncertainty or, in other words, the synergy among the knowledge functions. The results indicate that three regimes were generated in the Hungarian transition period with very different dynamics: (1) Budapest and its agglomeration emerge as a knowledge-based innovation system on every indicator; (2) in the north-western part of the country, foreign-owned companies and FDI have induced a shift in knowledge-organization; while (3) the system seems to be organized in the eastern and southern part of the country in accordance with government expenditures. The national level no longer adds to the synergy among these regional innovation systems.

## 1. Introduction

When the Soviet-Union fell apart in the years 1989-1991, the countries of Eastern and Middle Europe regained their autonomy. A major task which confronted national governments was to guide both the transition of their economies to a modern knowledge-based economy in rapidly globalizing markets and to achieve accession to the Western institutions such as NATO, the OECD, and thereafter also the European Union. The accession to the EU—to be realized in 2004—required a transformation of the economic and legal systems of these countries. The institutionally shaped bureaucracies of the socialist economies had not been able to absorb the emerging knowledge-based dynamics sufficiently (e.g., Richta *et al.*, 1968). The newly elected governments were confronted with both transformations by opening up their economies to the world market and the innovation dynamics of external forces such as transnational corporations. Foreign-driven investments (FDI) thus became decisive in shaping the national and regional innovation systems in these countries (INZELT, 2003, RADOSEVIC, 2002).

In this paper, we focus on Hungary as a country which went through this process of transition and thereafter accession to the EU. The study follows the format of previous studies of one of the authors about the measurement of regional and national systems of innovation in The Netherlands and Germany (LEYDESDORFF *et al.*, 2006; LEYDESDORFF and FRITSCH, 2006) and combines this statistical apparatus with the in-depth insights in the Hungarian system of the other author. The added value of this paper shall be the extension of the empirical model to transition economies. We have access to unique data that allow us to measure the effects of the transformation in terms of synergies in relevant innovation systems (CARLSSON, 2006). The data consists of micro-data of 660,290 firms in Hungary which are classified as high-tech or medium-tech manufacturing, and knowledge-intensive services by the Hungarian Central Statistical Office (HCSO). In addition to this classification, the data contains the postal addresses and the sizes of all these firms in terms of numbers of employees.

The second main contribution of this paper is a more refined theoretical background of the empirical model, in which we distinguished three knowledge functions of innovations systems, namely: knowledge exploration, knowledge exploitation, and organizational control. Following the before mentioned studies, we use size of the firm as an indicator of organization (PUGH et al, 1969, BLAU and SCHOENHERR, 1971), whereas the addresses allow us to decompose the regional dynamics at various levels of aggregation. Our research question is whether a national system of innovations was shaped in Hungary, and if not how the system should then be characterized in terms of differences among regions? We operationalize an innovation system as a configuration among technological, geographic, and organizational dimensions (STORPER, 1997), in which knowledge functions can generate synergy. As a methodology, we can then use configurational information as a measure of synergy among distributions in the three dimensions.

We shall argue that Hungary has now to be understood as composed of three innovation systems with very different dynamics: (1) the capital Budapest can be characterized as a metropolitan innovation system which operates increasingly on a par with Vienna, Prague, and Munich as centers for knowledge-intensive services and knowledge-based manufacturing; (2) the north-western parts of the country have been absorbed in the Western-European innovation systems which surround it (notably, Austria and Germany); and (3) the eastern and southern parts of the country are still predominantly integrated in terms of the old systems dynamics, that is, controlled by government spending. The national system of Hungary no longer adds to the knowledge-based dynamics of these three regional systems as it does in the case of the Netherlands. In Germany, we found synergy at the state level (which are also defined as NUTS-1). These conclusions will be quantitatively underpinned in terms of measuring synergies among the technological, organizational, and territorial distributions of firms at the various NUTS-levels.[1]

## 2. The evolutionary model

The literature on innovation systems is built upon the question why and how nations and regions differ in terms of the evolution in technology, industrial structure and institutional setting of certain territories (COOKE et al., 2004, EDQUIST, 1997, LUNDVALL, 1992, NELSON, 1993). The inter-relatedness of the above mentioned settings is discussed in institutional and evolutionary research agendas as the notion of co-evolution (BOSCHMA and FRENKEN, 2006, CORIAT and DOSI, 1998). However, the investigation of these co-evolutionary mechanisms remained mostly on the national level of innovation systems (NELSON, 1995). The model elaborated in this section enables us to investigate whether the inter-relatedness occurs on national or regional levels of innovation systems.

Our model is based on Storper's (1997, at pp. 26 ff.) conjecture that the interrelationships among technology, organization, and territory in an economic system can be considered as a "holy trinity." Storper emphasized that this holy trinity should not be studied as an aggregate of the composing elements, but in terms of the relations between and among these elements. These relationships shape regional economies. However, his proposal for a "relational paradigm" was not operationalized in terms which allow for measurement.

---

[1] NUTS is an abbreviation of Nomenclature des Unités Territoriales Statistiques. This classification was established by Eurostat more than 25 years ago in order to provide a single uniform breakdown of territorial units for the production of regional statistics for the European Union; at http://europa.eu.int/comm/eurostat/ramon/nuts/introduction_regions_en.html

Using the Triple Helix model of university-industry-government relations, Leydesdorff (2003) proposed configurational information as providing an indicator of synergy in Triple-Helix relations. This information measure—to be discussed in more detail below—can be negative or positive, indicating the existence of synergy among three independent sources of variance. For example, one can ask whether the interaction among these sources in a region is further enhanced by additionally considering the national level. By applying this concept to Storper's three dimensions, one can consider the following model. Here we used the organizational dimension of the "holy trinity" as operationalization of the economic exchange relations, as the latter can be expected to determine the size and scope of firms through transaction costs in the market (COASE, 1937).

Figure 1. Synergy of knowledge functions of an innovation system

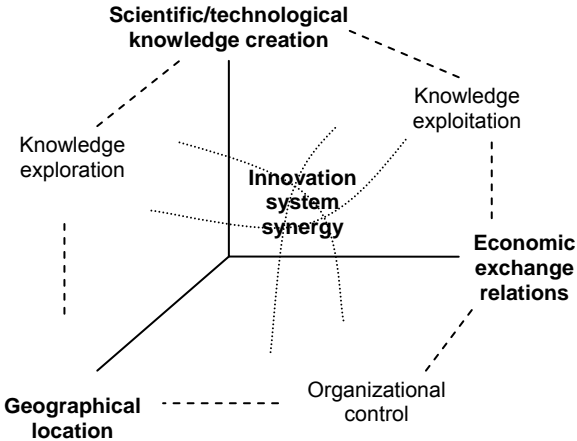

Source: adapted from LEYDESDORFF, 2006.

This model enables us to distinguish knowledge functions in innovation systems in addition to the two main dimensions of a political economy. Agents are geographically positioned and endowed, but are able to exchange irrespectively of these boundaries in economic relations. A knowledge-based economy can be considered as based on interactions between these two (traditional) drivers of a political economy with the new dimension of knowledge creation, dissemination, and control (NELSON and WINTER, 1982; WHITLEY, 1984, 2001). The three dimensions operating upon each other can also be considered as a triple helix which endogenously might be able to reduce uncertainty in a process of self-organization. As noted, our intent is to measure this reduction of uncertainty as a synergy, which will be explained in details below.

In our three-dimensional model knowledge functions of innovation systems are defined, namely: knowledge exploration, knowledge exploitation, and organizational control, which can be considered as the interaction terms at the interfaces between the three independent dimensions. This specification of the knowledge functions in terms of relevant dimensions enables us to connect the model to Storper's "holy trinity" beyond the operationalizations in the mentioned studies of innovation systems in the Netherlands and Germany.

*Knowledge exploitation* can be considered as the selection of existing routines, while *knowledge exploration* refers to concerted variation and planned experimentation (MARCH, 1991, BAUM et al., 2000). In other words exploitation is associated with the reuse of existing competences and means; while exploration is associated with creating new alternatives. At the systems level, mechanisms of knowledge exploitation represent the interface between economic welfare and technological knowledge creation (GIBBONS et al., 1994). In our understanding this interface does not necessarily depend on geographical locations, because economic welfare is created at the level of global markets, even if certain technologies originate in single regions. We also argue that knowledge exploitation is connected to locations only when synergy of innovation system exists: the local pool of suppliers, qualified labour etc. Knowledge exploration, however, can be considered as a place-dependent rather than market-dependent mechanism, because tacit knowledge is essential in creating new knowledge and it relates significantly to places. Strong evidence was found on the local effect of university R&D on economy (JAFFE et al., 1993) and on the regional production of knowledge (ACS et al., 2002, VARGA, 2007).

*Organizational control* mechanisms of innovation systems (e.g., economic policies) focus mainly on territorial units having direct influence on economic exchange relations (taxes and incentives etc.), but affect knowledge exploration and exploitation only indirectly (infrastructure, cluster-programmes etc.). Institutions of innovation systems increase the probability of the emergence of new knowledge; however, organizations are needed to foster it by organizational arrangements, new connections, and formal responsibilities (LOASBY, 2001). Thus, the *organizational control* function in our model represents the institutional and organizational elements of innovation systems. In our opinion, it follows from the above outlined knowledge functions that the synergies among them mark the quality of innovation systems.

As noted, this study leans methodologically on two previous studies for Germany and the Netherlands, respectively. The main conclusions of these studies provide the expectations in this study. These conclusions were that

- medium-tech and high-tech manufacturing couple knowledge synergies of innovation systems to geographical location;
- medium-tech does this to a higher extent than high-tech;
- knowledge-intensive services tend to uncouple the knowledge synergies from their geographical location;
- high-tech services counter-act on this latter effect.

The authors explained these findings in terms of the relative "footloose-ness" of knowledge-intensive services and high-tech manufacturing (VERNON, 1979). Knowledge-intensive services can be offered outside a region using means of transportation. Thus, the vicinity of an airport or major railway station is often more important than local and regional factors of the economy. High-tech manufacturing is less embedded in a local economy than medium-tech. The latter can be expected to enhance absorptive capacity in the region to an extent larger than the former (COHEN and LEVINTHAL, 1989). High-tech services counter-act on the uncoupling effect by making R&D facilities (e.g., laboratories) sometimes a necessary condition.

Furthermore, these authors found an additional synergy at the level of the nation in the case of Holland, but not in the case of Germany. In Germany, innovation systems are integrated more at the level of federal states than at the level of the nation according to this indicator. In the case of Hungary, we initially expected that Hungary would function like the Netherlands as a national system considering also the scale of the relatively small nation. As noted, our data will suggest that the Hungarian national system of innovations has in the meantime fallen apart in different subsystems, that is, parts of the country which operate with other dynamics. The coordination problem at the national level will also be discussed.

**3. Selected issues of the Hungarian transition: regional distribution of FDI and R&D expenditure**

Since the political changes in 1989 and 1990 the relatively small Hungarian economy has opened up. The country joined the EU in 2004 after a transition period. Articles in economic geography point out that transition from a planned into a market economy first caused a significant economic fallback (LENGYEL, 2004; VARGA, 2007): many medium-tech state-owned companies went bankrupted, R&D expenditure declined. Three regions (Central Hungary, Central Transdanubia, Western Transdanubia) actually began to reduce the gap between them and their Western-European counterparts with a growth of approximately 4-5% a year in the late '90s. These three regions with dynamically expanding economies constitute one block situated in the northwest of Hungary between Budapest and the Austrian border. Out of these three regions the Central Hungarian region (Budapest) has reached the level of 96.5 % of the EU-25 average by the year 2005 in terms of GDP per capita. The economic growth of the other four regions remained at a yearly 1.6-3% growth of GDP (HSCO, 2007). These regions (Southern Transdanubia, Northern Hungary, Northern Great Plain, Southern Great Plain) are situated south and east of Budapest. In 2003 the three developed regions produced 70% of Hungarian manufacturing exports, while the contribution of the Southern Great Plain region was only 6% (HSCO, 2004).

In the Hungarian economy more than 50% of the registered capital of companies and partnerships is in the hands of foreign owners. International stake is significant in all SME types, reaching at least 27-28 % (KÁLLAY and LENGYEL, 2008). Similarly to large companies, half of the capital of medium-sized enterprises is in foreign hands.[2] The R&D index in Hungary lags behind the European average: the total share of R&D expenditure of GDP was 0.95% in 2005; 44,8 % of the R&D expenditures were spent in the business sector, 29,1 % in public research organization, 26,1 % at universities (HCSO, 2006). Previous studies concluded that university-industry relations are weak in Hungary (INZELT, 2004; PAPANEK, 2000). However, some regional centres have important universities which have taken an active role in the transition. The Central-Hungarian Region (CHR)—including Budapest—plays a determining role in the Hungarian R&D performance (Figure 2).

---

[2] It is important to note that in Hungary an enterprise with at least 10 % of foreign business share falls in the definition of enterprises including foreign stake.

Figure 2. Distribution of foreign stake and R&D expenditure in Hungarian counties, 2005

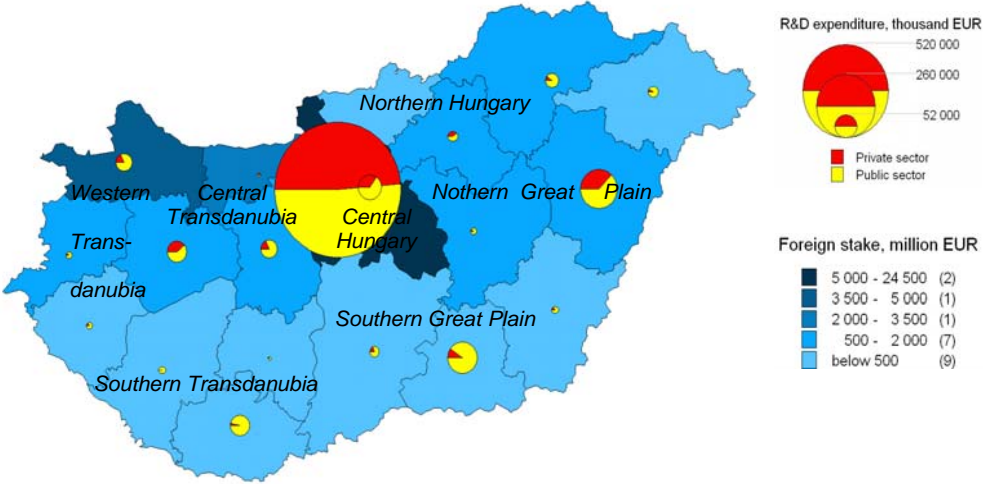

Source: Based on unique data reguest to HSCO and on data downloaded from the HSCO website: http://portal.ksh.hu/pls/ksh/docs/eng/xstadat/xstadat_annual/tabl6_03_02_05ie.html

The motives of foreign-owned companies for selecting specific locations (mainly in the energy-, banking and manufacturing sectors) were primarily the labour cost, accessibility, opportunities of privatisation in the beginning of the transition (BARTA, 2002). However, the knowledge base of Budapest and the regional centres seems to be more and more attractive for multinational R&D (LENGYEL and CADIL, 2008; INZELT, 2003); the growth of R&D spending by foreign affiliates was among the highests in Hungary in the period 1995-2003 (UNCTAD, 2005, p. 127). The share of large foreign-owned companies in manufacturing R&D is around 40 % (EUROSTAT, 2005); the share of foreign affiliates of business R&D is around 80% (EC, 2005). The territorial distribution of foreign investment in the Hungarian economy is uneven. The majority of foreign stake in foreign-owned companies are located in Budapest and Central Hungary, Central Transdanubia, and Western Transdanubia.

As noted, this paper focuses on the question how regional differences occurred during the transition by the entrance of foreign investors and the self-maintenance of public expenditures. We consider foreign direct investment as *external force* on innovation systems (LORENZEN and MAHNKE, 2002; LUNDVALL et al, 2002). Companies locate their R&D facilities to a region either to exploit their existing knowledge or to have access to the local knowledge base (VON ZEDTWITZ and GASSMAN, 2002). Public expenditures on research and development are transferred to the regions from the central budget and private R&D is mostly controlled by foreign-owned firms in Hungary. However, we follow here the general look on R&D as it is *internal mechanism* of the regional innovation systems (TÖDTLING and TRIPPL, 2005). In summary, we expect that

- foreign-owned firms have a restructuring effect on the synergy of knowledge functions;
- the contribution of research and development to the knowledge synergies is strongly differentiated by regional innovation systems.

**4. Data and methods**

In the analysis we used a unique dataset to measure the existence of synergy in different levels of innovation systems. Our indicator, the configurational information measures the reduction of uncertainty of three independent sources of variance in the system. The stronger is the synergy among knowledge functions the more uncertainty is reduced in the system. A three-dimensional matrix of data was needed to use this indicator, which accords to the model provided in Figure 1.

The dataset consists of 659,701 establishments and was collected by the Hungarian Central Statistical Office (HCSO). As it is obligatory for firms to supply data for the HCSO, our dataset can be considered as the entire population. The data applies to December 31, 2005. The use of this statistical register of enterprises provided us with the information at the company level: each company is classified into geographical, technological and organisational categories using the following systems.

The geographical dimension was investigated at the LAN 1 (previously called NUTS 4) level of sub-regions. Hungary as a whole is considered as a NUTS 1 unit according to the Eurostat classification. There are seven regions (NUTS 2), 19 counties (NUTS 3) and 168 subregions (LAN 1) in Hungary. Since the data was collected at the level of subregions, we are able to aggregate the information and define the uncertainty in geographical distributions at the NUTS 3 level of counties. Budapest is the only metropolitan district in Hungary and has hence to be considered as a special category in regional surveys: data from the Budapest is generally collected at the NUTS 3 level.

In order to measure the technological dimension we use the NACE code of industrial sectors developed by the OECD and Eurostat. Since various sectors of the economy can be expected to use different technologies, sector classifications can be used as a proxy for the technology (PAVITT, 1984). The OECD (2001, pp.137 ff.) indicated the various sectors in terms of their R&D intensity. Each enterprise is classified by its first activity at the two-digit level. Our data consists of all firms classified within one of the 22 categories listed in Table 1.

Table 1 around here

The dimension of economic exchange relations in our model is operationalized by organisational terms, namely the size of enterprises and measured by the number of their employees (PUGH et al., 1969; BLAU and SCHOENHERR, 1971). The Hungarian enterprise register has a category with zero or unknown employees that includes the SMEs without employee or self-employing. This category contains, among others, spin-off companies that are already on the market but whose owners are employed by mother companies or universities (Table 2). Surprisingly, a high percentage of firms (53.7%) is classified by the categories of high-tech, medium-tech industries and knowledge-intensive services.

Table 2 around here

We composed our dataset into a three-dimensional matrix of 168 lines (LAN 1 regions), 22 columns (NACE codes) and 6 levels (size categories).

*Methods*

The configurational information is closely connected to entropy measures. Entropy is widely used in geography as a measure of inequalities among or diversity in territorial units (BOSCHMA and IAMMARINO, 2007). We apply it as a measure of uncertainty contained in a probabilistic distribution or system of distributions (Johnston *et al.*, 2000). According to SHANNON's (1948) formula,[3] the uncertainty in the distribution of variable x ($\sum_x p_x$) is defined as $H_x = - \sum_x p_x \log_2 p_x$. Analogously, for two dimensional distribution $H_{xy} = - \sum_x \sum_y p_{xy} \log_2 p_{xy}$. This uncertainty is the sum of the uncertainty in the two dimensions diminished by their co-variation.

---

[3] The higher is the indicator of $H_x$ the more diverse is the system, thus the higher is the level of uncertainty that prevails.

When the basis of the logarithm is two, the values are expressed in bits of information. Therefore our entropy measures are formal (probability) measures and thus independent of size or any other reference to the empirical systems under study. The sigma in the formula allows all the information terms to be fully decomposed. Our analysis measuring how the system effects the decrease in uncertainty is built on these characteristics of the entropy.

In the case of two dimensions, the uncertainty in the two potentially interacting dimensions (x and y) can be reduced with the common entropy. Our aim is to mark systems, thus we use the concept of configurational information or transmission, which captures the reduction of uncertainty and is formalized in two dimensions as follows:

$$T_{xy} = (H_x + H_y) - H_{xy} \qquad (1)$$

In the limiting case that the distributions x and y are completely independent, $T_{xy} = 0$ and $H_{xy} = H_x + H_y$. In all other cases $T_{xy} > 0$, and therefore $H_{xy} < H_x + H_y$ (THEIL, 1972, at pp. 59f.). In general, two interacting systems (or variables) determine each other in their mutual entropy ($H_{xy}$). However, in the case of three interacting variables, one has two options: the three interacting systems may have a common segment shared by all of them or not (Figure 3a and 3b, respectively). In the latter case, the overlap can also be considered as negative.

Figure 3: Relations between probabilistic entropies (*H*), transmissions (*T*), and configurational information for three interacting variables.

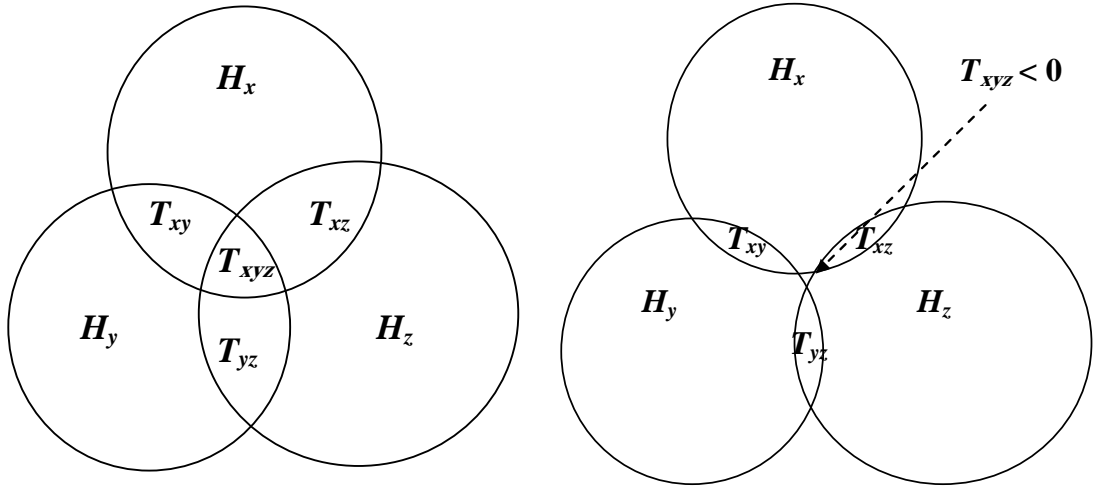

In the case of overlaps, the mutual informations are redundant, but in the other case the cycling of the information between each two dimensions can generate a synergy. This redundancy of synergy can be expressed by an information measure ($T_{xyz}$) which ABRAMSON (1963, at p. 129) derived from the Shannon formulas:[4]

$$T_{xyz} = H_x + H_y + H_z - H_{xy} - H_{xz} - H_{yz} + H_{xyz} \qquad (2)$$

---

[4] Both YEUNG (2008, p. 59f.) and KRIPPENDORFF (2009, p. 200) noted that this information measure can no longer be considered as a Shannon-type measure because of the possible circularity in the information transfers. Shannon-type entropy measures are by definition linear and positive. Since the measure sums Shannon-type measures in terms of bits of information, its dimensionality is also bits of information, and therefore it can be used as a measure of uncertainty and uncertainty reduction, respectively. YEUNG (2008, at pp. 51 ff.) further formalized the configurational information in three or more dimensions into the I-measure $\mu^*$.

While the two-dimensional systems reduce the uncertainty, the trilateral term in turn feeds back on this reduction, and therefore adds another term to the uncertainty. Thus, the configuration of the system determines the net result in terms of the value of $T_{xyz}$ (MCGILL, 1954).

As noted, the three dimensions under study in this case will be (G)eography, (T)echnology, and (O)rganization, and the configurational information among them will accordingly be indicated as the $T_{GTO}$. Similarly to Equation 2 one can formulate as follows:

$$T_{GTO} = H_G + H_T + H_O - H_{GT} - H_{GO} - H_{TO} + H_{GTO} \qquad (3)$$

The value of $T_{GTO}$ measures the interrelatedness of the three sources of variance and the fit of the relations between and among them. The synergy among knowledge exploitation, knowledge exploration and organizational control reduces the uncertainty in the system. We use $T_{GTO}$ as a measure of the reduction of the uncertainty at a systems level: a better fit will be indicated with a more negative value. Note that the indicator does not measure the innovative activity or economic output of a system. It measures only the structural conditions in the system for innovative activities, and thus specifies an expectation.

**5. Results**

As noted, the data allow us to disaggregate in terms of geographical regions (NUTS 2 and NUTS 3), and we are able to distinguish high-tech and medium-tech sectors versus knowledge-intensive services. The various dimensions can also be combined in order to compute the configurational information in the interaction among them in a next step.

The counties are different in terms of the numbers of firms and their geographical distributions inside the counties (Table 3). While according to NACE categories only 8,722 enterprises proved to be high-, medium-tech or knowledge-intensive in Nógrád county, Budapest contains 229,165 firms, and the Pest county 67,342 enterprises. The mean of analysed firms at the county level (without Budapest) is 22,690.

*5.1. Entropy values*

As the data were collected at the LAN 1 level, the first column of Table 3 informs us about the uncertainty of the sub-regional distribution at the county level or, in other words, about the concentration of economic activities. Budapest is a special case, as it is at the same time a NUTS 3 and a LAN 1 region. Thus, there can be no uncertainty for the concentration within the Budapest area. However, the Pest county has a large value on this indicator, it contains 15 subregions. In this case, 3.54 bits of information is equal to 90.7% of the maximum entropy of $\log_2(15) = 3.91$. One can understand this as a representation of the spread of economic activities in more sub-regions of the Budapest agglomeration. The other counties with a high number of sub-regions are more centralised. For example, the distribution of firms in Borsod-Abaúj-Zemplén (with Miskolc as the second largest settlement in Hungary) corresponds to 62.7% of the maximum entropy. The counties with strong university centres are also more centralised; in these cases the value of probabilistic entropy shows a lower percentage of the maximum entropy: Csongrád 62,5%, Baranya 54,2%, Hajdú-Bihar 59%, Győr-Moson-Sopron 66,2%.

Table 3 around here

The maximum entropy in the technological distribution is $\log_2(22) = 4.459$, and $\log_2(6) = 2.584$ in the organisational distribution. The entropy values in both dimensions indicate a very skewed distribution on the country level. In the technological and organisational dimensions the percentages of this maximum entropy are 61.1% and 44.9% at the level of the nation. In the analyses the probabilistic entropy indicates the uncertainty only among the high- and medium-tech industries and knowledge-intensive services.

The probabilistic entropies ($H_t$ and $H_o$) in the county decomposition show a relatively small variance; counties do not differ in terms of their organizational or technological variety. The value of H for the country as a whole corresponds to the mean of the values for counties in technological and organisational dimensions: $H_t = 2.744 \pm 0.071$ and $H_o = 1.145 \pm 0.035$ bits of information. The low percentages of expected entropy indicate asymmetric distributions in the organisational dimension. A possible reason for these skewed distributions could be the large share of micro companies and firms without employees in our data.

Surprisingly, 75% of the total number of registered firms without employees in Hungary is represented in the 22 NACE categories of high- and medium-tech industries and knowledge-intensive services (see Table 2). The category of firms without employees in the real-estate sector (NACE 70) is a good example for illustrating this strongly uneven distribution. The number of them is 140,078; this category contains 21,2% of all the establishments included in the analysis. The category of micro firms with other business activities (NACE 74) has a similar weight: 156,807 units (23.7%).

The entropy values in two dimensions reduce the uncertainty of the system and can be used as proxies to measure the knowledge functions outlined in section 2. Consequently, $H_{gt}$ is a proxy for knowledge exploration, $H_{to}$ is a proxy for knowledge exploitation and $H_{go}$ is a proxy for organizational control. However, one can observe that redundancies in the geographical and technological dimensions ($H_{gt}$), are relatively low in the counties with big universities (Baranya, Csongrád, Győr-Moson-Sopron, Hajdú-Bihar). Consequently, the interdependence among the geographical and technological distribution of the studied industries is low in these areas. On the other hand, these interdependences seem to be high in the Pest county, though it has more centres, more universities, and therefore a more diversified economy. This means that knowledge exploration occurs on a higher level in the broader Budapest agglomeration than in university towns, where one could also expect knowledge exploration to stand out. It is an interesting issue for further research to prove this statement. We focus our attention on the configurational information in three dimensions and analyze the main anomalies at the systems level as follows.

*5.2. The geographical decomposition of the configurational information*

While high values for the mutual information between two dimensions indicate the strength of interaction among them, the configurational information of three interacting dimensions can be negative or positive. The value of configurational information depends on the relation among the entropy values of two-dimensional and three-dimensional distributions (see Equation 3 above). The question in our case is, whether the knowledge functions of the innovation system reduce the uncertainty among the geographical, technological and organizational distributions. The synergy among knowledge exploration, knowledge exploitation and organizational control may reduce the three-dimensional term (YEUNG, 2008). Thus, a more negative value of the configurational information in three dimensions indicates a decrease in the uncertainty prevailing at the systems level.

The values for the configurational information in three dimensions are negative for all the counties (Table 4). This indicates that counties are relevant innovation systems in Hungary.

Table 4 around here

The values for $T_{GTO}$ already take account for the geographical distributions as one of the relevant dimensions. Since the number of regions varies among counties, these values cannot directly be compared with one another. However, it can be noted that the value of $T_{GTO}$ is less pronounced for Hungary as a country than for any of its parts. In order to make this comparison among regions possible, we weighted the values of $T_{GTO}$ with the number of firms in the counties. ΔT in millibits represents these values that we can interpret as the measure of innovation systems synergies. There is an inverse relation between the absolute value of the indicator and the reduction of uncertainty in the system.

Our line of argument calls for marking the national innovation system first. In order to do this we have to use the feature of entropy statistics that the values can be fully decomposed (THEIL, 1972):

$$T = T_0 + \sum_i n_i/N \times T_i \qquad ; \qquad (4)$$

where $T_0$ is the in-between county entropy,
$T_i$ is the entropy measured in county i,
$n_i$ is the number of firms in county i,
N is the number of firms in the whole country.

The in-between group uncertainty ($T_0$) is defined as the difference between the sum of the uncertainties of the contributions and the uncertainty prevailing at the level of the composed set (LEYDESDORFF et al., 2006). In this case the $T_0$ is an indicator of the in-between group contribution to the configurational information in three dimensions. A negative value would indicate that the national agglomeration adds to the synergy in the system, while a positive value indicates that the synergy occurs more on regional than national levels.

The high *positive* value of $T_0$ indicates that Hungary is far from integrated as a national innovation system. The in-between region term adds to the uncertainty at the national systems level. Previous studies found a negative value for the Netherlands (LEYDESDORFF et al., 2006) and separate German states, but a positive one for Germany as a nation (LEYDESDORFF and FRITSCH, 2006). Unlike the Netherlands, Germany is both a federal state and a country with some eastern states "in transition." Thus, one would not expect this synergy at the national level. The results for Hungary, however, are more dramatic because Hungary is a nation-state and the contribution of the in-between county uncertainty ($T_0$ = +10.94 mbits) is far larger than any of the reductions of the uncertainty at the regional level. We will elaborate our further tests on this finding and investigate the forces that shaped innovation systems in the Hungarian transition.

Figure 4: The configurational information among three dimensions at NUTS 3 level in Hungary

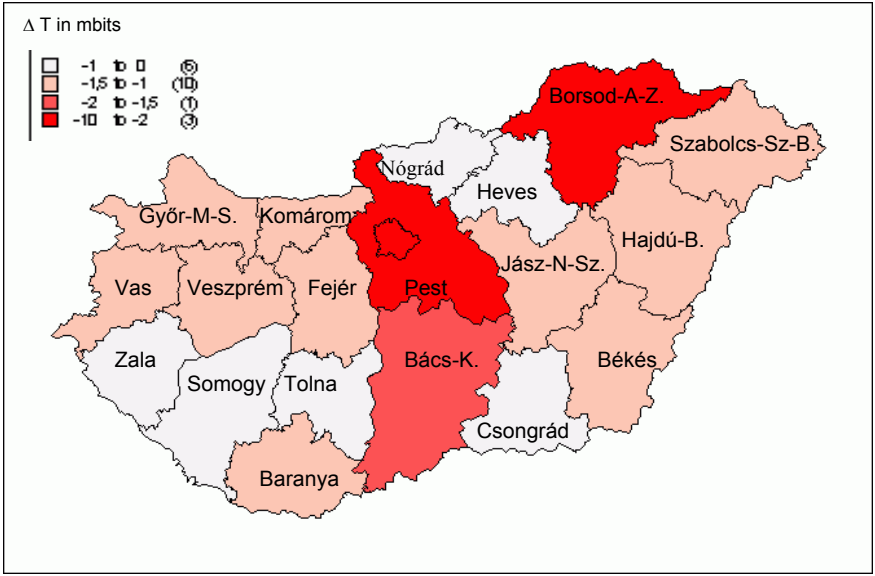

The synergies among knowledge functions of Hungary stand out in Budapest (ΔT =−9.63 mbits), the Pest county (−3.39 mbits), and Borsod-Abaúj-Zemplén (−2.39 mbits). In other words, our results suggest that the strong differentiation in terms of regional prosperity can be verified using this methodology: the knowledge base of Budapest and its agglomeration becomes visible as central to the country's economy (Figure 4).

In addition to the strong position of Central Hungary a few annoying problems emerge. While one can observe parallel ranking among counties in the West in terms of configurational information and indicators of economic performance or internalization, one does not find such correspondences in the eastern part. For example, the high level of integration in Borsod-Abaúj-Zemplén is unexpected: the employment rate in that area in the North-East is among the lowests and the economic performance is the lowest in Hungary. It is also surprising that the value of configurational information in Csongrád and Hajdú-Bihar do not emerge, though these are locations of relatively large universities.

Most of the studies that analyzed the transition period from a deterministic view suggested that Budapest and the north-western part of Hungary where most of the multinational firms have located their sites (we introduced these trends in section 3) have developed a strong endogenous knowledge base (ANTALÓCZY and SASS, 2005; INZELT, 2003; SZALAVETZ, 2004; TÖRÖK and PETZ, 1999). As our analysis is based on complexity perspectives and on stochastic measures, our results indicate otherwise. Might it be that the foreign-owned corporations have disturbed the self-organization of local innovation systems, the texture of university-industry-government relations in which medium-tech companies can play a leading role in integrating the system?

We suggest that with the exception of the metropolitan area of Budapest, the foreign-owned capital uncoupled the knowledge synergies from its geographical rooting. These companies have their headquarters in Western capitals, decisions are not made in the local systems. In the terms o knowledge functions we argue that exploitation came to the front after the open up of new markets, which also contributed to the process that local synergies transformed in. Thus, the dynamics are very different from those of the Netherlands and (Western-) Germany because these latter systems did not go through a transition period. Could this geographical uncoupling mean that the areas in the West are integrated in the EU or neighbouring countries more than in the national system? Previous empirical studies found that foreign direct investment from Austria is concentrated in the North-Western regions of Hungary, close to the Austrian border (INZELT, 2003, at p. 256). Moreover, these regions are also included in the Regional Innovation Strategy of Vienna, the capital of Austria (BORSI et al., 2007). The eastern and southern part of Hungary is least affected by the transition hitherto and therefore innovation system synergies are more in place.

## 6. Investigating our expectations

While the geographical comparison is compounded with traditional industrial structure like firm density, all effects of the decomposition in terms of the sectoral classification of high- and medium-tech sectors and knowledge-intensive services can be expressed as relative effects, that is, as percentages increase or decrease of the negative value of the configurational information in three dimensions when a specific selection is compared with the population. In the remainder of this study, we use the categories provided by the OECD and Eurostat (see Table 1 above) as selection criteria for subsets and compare the results with those of the full set—provided in the previous section—as a baseline. The sturcure of this section develops the line of our expectations: the first subsection reflects on the findings of previous studies in the Netherlands and Germany, while the second and third subsections investigate the expectations in Hungary as a transition economy.

*6.1. The role of geographical location in the synergy among knowledge functions*

The investigation of our expectations need first a comparison of the sectoral effects on configurational information in three dimensions. The sectoral effects were calculated as shares in the configurational information of the total system:

$$(\text{Sector } T_{gto} - \text{Total } \Delta T) / \text{Total } \Delta T \times 100 \qquad (5)$$

In this case, a negative value means that uncertainty reduction is lower in the sector than the reduction in the total set, and a positive value indicates strong sectoral effect on knowledge function synergy in the region than on average.

The number of companies in high-tech and medium-tech sectors is vanishing compared to knowledge intensive services but these sectors have strong effects on the knowledge function synergy in the innovation system (Table 5). This small subgroup has an enormous effect on the configurational information in three dimensions, thus on the reduction of uncertainty; in some cases even more than 1000%. The selection of high- and medium-tech manufacturing makes large differences. Budapest and Pest are emerging considering the strength of synergy in these sectors followed by counties like Fejér, Komárom-Esztergom, Borsod-Abaúj-Zemplén, Jász-Nagykun-Szolnok, and Bács-Kiskun (column 7 in table 5). The knowledge function synergies are the weakest in South Transdanubia and Southern Great Plain.

Table 5 around here

One can observe that high-tech and medium-tech companies had positive effects on the reduction of uncertainty in all counties and at the national level as well. Knowledge exploration, knowledge exploitation, and organizational control have synergic effects on each other in these sectors.

Comparing the sectoral effects of knowledge-intensive services on the configurational information our findings are similar to the results of studies following similar methods in the Netherlands and Germany (LEYDESDORFF et al., 2006; LEYDESDORFF and FRITSCH, 2006). Column 4 of Table 5 indicates that the knowledge-intensive services reduce uncertainty less than the total set in all regions (the rates are all negative). We may therefore conclude that knowledge-intensive services (KIS) have negative effects on the knowledge function synergy in Hungarian regional innovation systems. These services can be provided in a bigger distance; consequently they may link the local self-organization of regional innovation systems into global contexts. The negative effect seems less pronounced than in the Netherlands or Germany, but this is an artefact of the comparatively large size of this subset in the population of firms: 97.1% of the units of analysis are classified in this category. We may assume that these include also less active firms.

In summary, we found that medium-tech and high-tech manufacturing couple knowledge synergies of innovation systems into geographical location in Hungary. Knowledge-intensive services tend to uncouple the knowledge synergies from the geographical location. These findings are similar to the results in previous studies in The Netherlands and Germany.

*6.2. The effect of foreign-owned companies*

Figure 5 shows the reduction of uncertainty because of configurational information in high-tech and medium-tech sectors across the various regions. The selection of high- and medium-tech manufacturing makes large differences; the effects of Budapest agglomeration stand out. One can find an emerging North-South pole in the eastern part of the country.

Figure 5: The configurational information considering the high- and medium-tech sectors at NUTS 3 level in Hungary

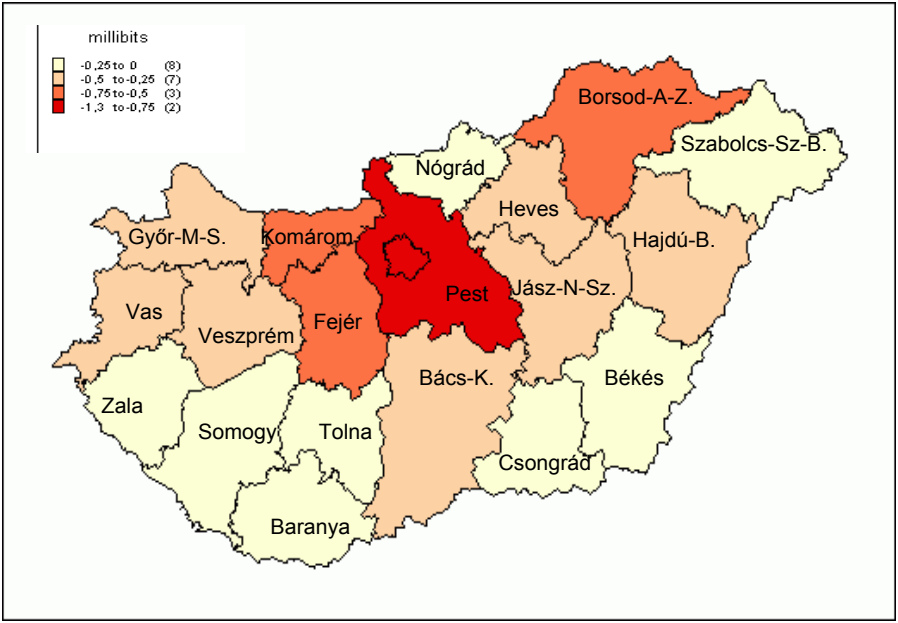

Not Budapest, but Fejér, Komárom-Esztergom, Pest, Jász-Nagykun-Szolnok, and Vas show the largest effects (column 5 in Table 5). The knowledge function synergy in these counties appears to be driven by mostly foreign-owned high-tech and medium-tech companies. Unfortunately we could not deal with company ownership as a fourth dimension of our model, likewise had no longitudinal data to analyze the relation between foreign investments and configurational information. However we can highlight some interesting correspondences with the results of a spatial analysis (Figure 6).

Figure 6. Foreign stake in foreign-owned companies and the knowledge function synergy in high-tech and medium-tech industries, 2005

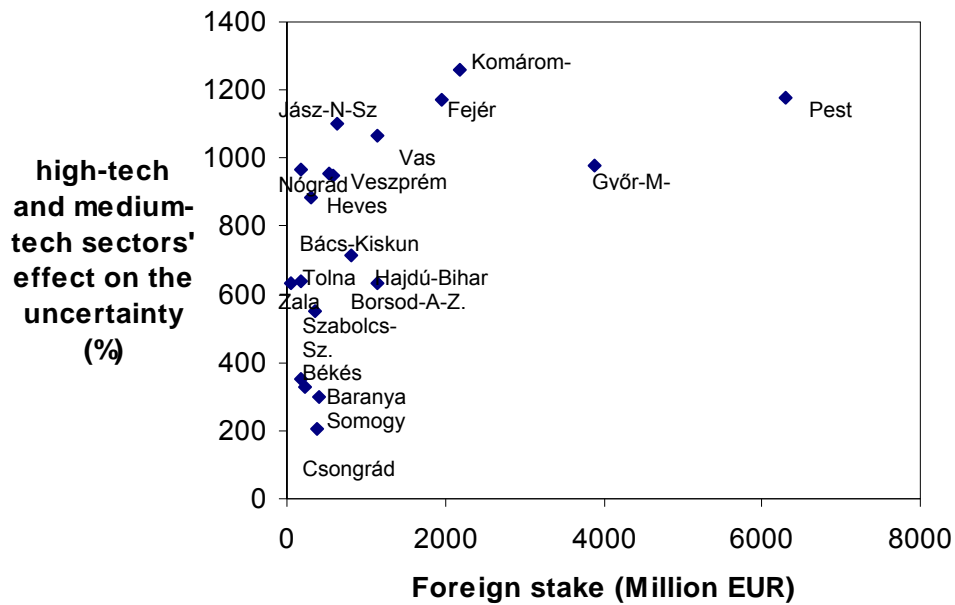

In Figure 6, we plotted the values of high-tech and medium-tech industries' effect on entropy against the county measures of foreign stake in foreign owned companies. The correlation between the two indicators (0.53) becomes significant (at the 5% level) only when we take out Budapest from the sample. Consequently, different tendencies prevail in the capital and in the counties. One can observe a moderate positive relation among the foreign investments and knowledge function synergies of high-tech and medium-tech industries, which is due to the decisive role of foreign-owned firms in both knowledge exploration (e.g. their major part in business R&D) and the open-up in knowledge exploitation.

Foreign stake in foreign-owned companies reached 24 500 million EUR in Budapest, exceeds two billion EUR only in four other counties: Pest, Komárom-Esztergom, Győr-Moson-Sopron, and Fejér. In these counties the high-tech and medium-tech sectors contribute importantly to the synergy among knowledge functions in the system. The results suggest that the rise of value added and the development of supply chains caused the positive effects during the transition. However, it must not to be forgotten that these sectors contain only a low number of companies that can change seat fast; their stake can alter even faster. In other counties one can observe the low level of foreign investment to be linked to various levels of knowledge function synergy.

This result accords with the expectation concerning the restructuring effect of the entrance of foreign-owned companies on innovation systems. Foreign-owned firms played a significant role in the transition of innovation systems in Hungary, through the privatization of R&D facilities, green-field investments in knowledge exploration and exploitation. However, a conditional statement has to be made here: foreign investments induce different spatial synergy among knowledge exploitation, knowledge exploration, and organizational control; Western regions emerge in this sense from the country.

*6.3. Effects of high-tech knowledge-intensive services and R&D*

High-tech knowledge-intensive services (HT-KIS), that have a core role in investigating our last expectation, are only a minor part of KIS: research and development, IT services, and post- and telecommunication services belong to this category. In our opinion, the effects of HT-KIS and the remaining part of KIS have to be compared as HT-KIS can mean knowledge- and technology creation whereas KIS can be considered as representative for the adaptation of knowledge and technology. More than knowledge-intensive services in general, high-tech services can be expected to produce and transfer technology-related knowledge (BILDERBEEK et al., 1998) and R&D is also expected to have stronger local knowledge spillover effects (VARGA, 2007) than KIS. However, we find a relative decline of the configurational information in three dimensions in most regions when the subset of sectors indicated as 'high-tech services' is compared with KIS (Table 6).

Table 6 around here

One can raise the question, to what extent are the trends of local knowledge transfer similar in the three sectors of HT-KIS? According to Miozzo and Soete (2001) telecommunication- and IT services belong to information network services while research and development belongs to science based services. Their taxonomy implies that the former group of services has an increased transportability and "making these networks work up to full capacity" assimilate them with scale-intensive services. From this perspective, post- and telecommunications, and IT services have different effects on knowledge function synergies than R&D. The former rather uncouple them from local circumstances, while the later couples them to spatial systems, like it was found in the case of some East-German regions (LEYDESDORFF and FRITSCH, 2006). According to our dataset post- and telecommunication, and IT services are more spatially distributed than R&D, which is more concentrated by nature. For example, from the seven R&D organization with more than 250 employees five are located in Budapest, two in Szeged, the seat of Csongrád county.

Figure 7: Contribution of high-tech services to the configurational information in three dimensions (% of normalized ΔT in mbits by HTKIS)

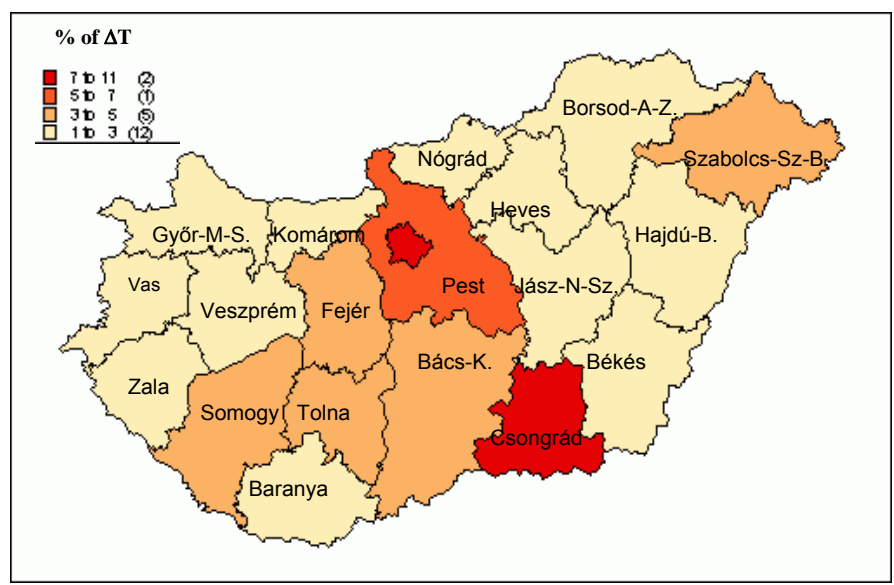

The HT-KIS effects of strengthening the system synergy appear only in Budapest and Csongrád (Figure 7), they have also the highest values of public research expenditures. However, one can argue that the reasons for these effects are different: Budapest has a strengthening knowledge base in the business sphere and the employment in high-tech knowledge-intensive services is higher than 4.5 % of the total employment (EUROSTAT, 2007, at p. 5). Our data shows that Csongrád is relatively strong in basic research (see Figure 1), but the economic sphere is relatively weak (LENGYEL, 2009). Thus, knowledge function synergy in HT-KIS means different mechanisms in these areas. The coupling effect of R&D is stronger than the uncoupling effect of post- and telecommunications and IT services in Csongrád; while large centres of these services integrate the innovation system synergies in the capital.

In summary, the contribution of research and development to the knowledge synergies are strongly differentiated by regional innovation systems. Research and development seem to couple knowledge function synergy of HT-KIS to the geographical location in two emerging centres, while post- and telecommunications and IT services might uncouple the synergy from spatial circumstances in other regions stronger than the coupling effect of R&D. Public R&D still plays a major role in raising the knowledge synergies in the south-eastern part of the country. This means that transition in the Eastern part of the country, where foreign investments are moderate, has been dependent on the institutions of the state, represented by organizational control in our model in section 2.

## 7. Conclusion and discussion

The analysis showed that high-tech and medium-tech manufacturing strengthens the geographical characteristics of the knowledge-based economy in Hungary. In all cases they have a reducing effect on the uncertainty in regional innovation systems. Knowledge-intensive services have weaker effects in uncoupling from the geographical dimension in Hungary than it was found in the Netherlands and Germany (LEYDESDORFF et al., 2006, LEYDESDORFF and FRITSCH, 2006). High-tech knowledge-intensive services couple the knowledge functions synergy to geographical locations in centres of R&D, like it was found in former East-German cases as well. The values of configurational information based on the high- and medium-tech industries are more pronounced in the regions where foreign-owned firms have a higher share.

One has to consider that foreign direct investments in Hungary were initially led by low labor costs. These trends have been followed after a 10-15 years acclimatisation period by investments to reach research potentials; this happened for example in the case of Audi in North-West Hungary (LENGYEL et al., 2006). Audi is one of the leading companies in terms of share of the Hungarian GDP production and also the leading role in the local automotive cluster, with an extended supply chain, its own institute at the local university, etc. Thus, these companies transform knowledge exploration and knowledge exploitation of innovation systems. Building on this example, one could argue that the synergy of three knowledge functions in high- and medium-tech industries measured by the configurational information is strongly reconstructed by foreign direct investments in Hungary.

According to these results we could identify three archetypes or paths of regional innovation systems in terms of knowledge function synergy: Budapest as an agglomeration economy; the north-western part of the country, which integrated to the European Union; the southern and eastern part of the country, where central planning and public R&D still have major effect on the innovation system synergy.

When Hungary entered its transition period it was probably too late to shape the National Innovation System with innovation policy initiatives to fulfil the requirements of global competition (ENYEDI, 1995). In the period of transition from a centrally planned economy towards the market economy and under the pressure of globalization and Europeanization, the Hungarian system was restructured not only in terms of linkages within the production system, but also in relation to its relevant environments. During this process the internal linkages were weakened and external linkages asynchronously reinforced (BARTA, 2002). Budapest and the north-western part of the country could find a way to the European market more easily than the eastern part. Universities could further develop international relations which had never been ceased to exist completely, and FDI became a major factor in the transformation process. Thus, the Hungarian system may have lost control over its political economy to an extent larger than traditional economies like the Netherlands which have been able to transform and adapt their national structures more gradually.

Our analysis was based on a three-dimensional model of innovation system synergy built on Storpers' 'holy trinity'. The indicator, the measurement of reduction in the uncertainty, gives the implication for further theoretical and empirical research. Namely, the complexity aspects of evolutionary economic geography raise the question how external forces act in local systems (MARTIN and SUNLEY, 2007). We believe that such forces, public R&D spending and foreign investments, might induce a shift in the uncertainty through bifurcation effects, while uncertainty is reduced in the system when agents develop their expectations of others. Complexity aspects also highlight that agglomeration trends lead to far from equilibrium states, in which the measurement of uncertainty in innovation systems might open up new ways of research. Our model explained in section 2 also needs further elaboration. One might find it interesting to make knowledge functions itself ready for empirical testing and underpin Storpers' 'holy trinity' with these terms. However, these arguments remain for further papers.

Table 1: Classification of high-tech and knowledge-intensive sectors

| High-tech Manufacturing | Knowledge-intensive Services (KIS) |
|---|---|
| **30** Manufacturing of office machinery and computers<br>**32** Manufacturing of radio, television and communication equipment and apparatus<br>**33** Manufacturing of medical precision and optical instruments, watches and clocks<br>*Medium-high-tech Manufacturing*<br>**24** Manufacture of chemicals and chemical products<br>**29** Manufacture of machinery and equipment n.e.c.<br>**31** Manufacture of electrical machinery and apparatus n.e.c.<br>**34** Manufacture of motor vehicles, trailers and semi-trailers<br>**35** Manufacturing of other transport equipment | **61** Water transport<br>**62** Air transport<br>**64** Post and telecommunications<br>**65** Financial intermediation, except insurance and pension funding<br>**66** Insurance and pension funding, except compulsory social security<br>**67** Activities auxiliary to financial intermediation<br>**70** Real estate activities<br>**71** Renting of machinery and equipment without operator and of personal and household goods<br>**72** Computer and related activities<br>**73** Research and development<br>**74** Other business activities<br>**80** Education<br>**85** Health and social work<br>**92** Recreational, cultural and sporting activities<br><br>Of these sectors, **64, 72** and **73** are considered *high-tech services*. |

*Source*: Laafia, 2002: 7.

Table 2: Distribution of company data by size

| **Number of employees** | **Number of firms included in this study** | **Number of registered firms – 31st Dec. 2005** | **% included** |
|---|---|---|---|
| 0 or unknown | 275,202 | 365,861 | 75 |
| 1-9 | 369,280 | 805,209 | 46 |
| 10-19 | 5,976 | 20,870 | 29 |
| 20-49 | 4,921 | 11,046 | 45 |
| 50-249 | 3,733 | 4,860 | 77 |
| 250 or more | 589 | 944 | 62 |
| Total | 659,701 | 1,228,999 | 54 |

Source: Hungarian Central Statistical Office (HCSO)

Table 3: Values of probabilistic entropy of the distributions in three dimensions and their combinations, bits

| Regions | Counties | $H_{geography}$ | $H_{technology}$ | $H_{organisation}$ | $H_{gt}$ | $H_{go}$ | $H_{to}$ | $H_{gto}$ | Number of firms | Number of subregions |
|---|---|---|---|---|---|---|---|---|---|---|
| Central Hungarian Region | Budapest | 0.000 | 2.598 | 1.169 | 2.598 | 1.169 | 3.644 | 3.616 | 229,165 | 1 |
| | Pest | 3.544 | 2.786 | 1.120 | 6.311 | 4.662 | 3.755 | 7.245 | 67,342 | 15 |
| Western Transdanubia | Győr-Moson-Sopron. | 1.858 | 2.658 | 1.130 | 4.500 | 2.985 | 3.577 | 5.380 | 28,177 | 7 |
| | Vas | 2.061 | 2.711 | 1.172 | 4.750 | 3.225 | 3.674 | 5.655 | 14,490 | 9 |
| | Zala | 1.978 | 2.717 | 1.155 | 4.679 | 3.132 | 3.663 | 5.595 | 16,538 | 6 |
| Central Transdanubia | Fejér | 2.345 | 2.715 | 1.152 | 5.043 | 3.493 | 3.701 | 5.984 | 24,075 | 10 |
| | Komárom-Esztergom. | 2.496 | 2.747 | 1.185 | 5.229 | 3.679 | 3.700 | 6.131 | 17,760 | 7 |
| | Veszprém | 2.739 | 2.756 | 1.144 | 5.474 | 3.880 | 3.671 | 6.342 | 20,533 | 9 |
| Southern Transdanubia | Baranya | 1.717 | 2.790 | 1.139 | 4.483 | 2.853 | 3.742 | 5.402 | 25,308 | 9 |
| | Somogy | 2.445 | 2.804 | 1.160 | 5.218 | 3.601 | 3.767 | 6.135 | 15,680 | 10 |
| | Tolna | 2.084 | 2.699 | 1.122 | 4.761 | 3.203 | 3.652 | 5.677 | 12,344 | 5 |
| Northern Hungary | Borsod-Abaúj-Zemplén | 2.449 | 2.809 | 1.138 | 5.238 | 3.584 | 3.769 | 6.142 | 30,174 | 15 |
| | Heves | 2.174 | 2.832 | 1.195 | 4.991 | 3.366 | 3.788 | 5.901 | 15,095 | 7 |
| | Nógrád | 2.225 | 2.771 | 1.186 | 4.982 | 3.405 | 3.687 | 5.841 | 8,722 | 6 |
| Northern Great Plain | Hajdú-Bihar | 1.871 | 2.743 | 1.130 | 4.596 | 2.998 | 3.687 | 5.505 | 26,624 | 9 |
| | Jász-Nagykun-Szolnok | 2.215 | 2.801 | 1.181 | 4.996 | 3.392 | 3.772 | 5.920 | 16,513 | 7 |
| | Szabolcs-Szatmár-Bereg | 2.435 | 2.842 | 1.116 | 5.251 | 3.548 | 3.792 | 6.158 | 20,422 | 11 |
| Southern Great Plain | Bács-Kiskun | 2.574 | 2.769 | 1.174 | 5.329 | 3.745 | 3.742 | 6.258 | 25,158 | 10 |
| | Békés | 2.678 | 2.574 | 1.067 | 5.189 | 3.733 | 3.537 | 6.096 | 19,003 | 8 |
| | Csongrád | 1.755 | 2.767 | 1.067 | 4.506 | 2.819 | 3.686 | 5.397 | 26,122 | 7 |
| Hungary | | 5.189 | 2.722 | 1.159 | 7.875 | 6.334 | 3.712 | 8.823 | 659,701 | 168 |

Table 4: The mutual information in three dimensions statistically decomposed at NUTS 3 level (counties) in millibits of information

| Regions | Counties | $T_{tgo}$ in millibits | $\Delta T$ in millibits | Number of firms |
|---|---|---|---|---|
| Central Hungary | Budapest | -27.75 | -9.63 | 229,165 |
| | Pest | -33.22 | -3.39 | 67,342 |
| Western Transdanubia | Győr-Moson-Sopron | -34.13 | -1.46 | 28,177 |
| | Vas | -48.89 | -1.07 | 14,490 |
| | Zala | -27.78 | -0.70 | 16,538 |
| Central Transdanubia | Fejér | -39.93 | -1.46 | 24,075 |
| | Komárom-Esztergom | -49.70 | -1.34 | 17,760 |
| | Veszprém | -43.45 | -1.35 | 20,533 |
| Southern Transdanubia | Baranya | -29.59 | -1.13 | 25,308 |
| | Somogy | -41.87 | -0.99 | 15,680 |
| | Tolna | -33.95 | -0.63 | 12,344 |
| Northern Hungary | Borsod-Abaúj-Zemplén | -52.32 | -2.39 | 30,174 |
| | Heves | -42.19 | -0.96 | 15,095 |
| | Nógrád | -50.37 | -0.67 | 8,722 |
| Northern Great Plain | Hajdú-Bihar | -31.93 | -1.29 | 26,624 |
| | Jász-Nagykun-Szolnok | -42.04 | -1.05 | 16,513 |
| | Szabolcs-Szatmár-Bereg | -38.53 | -1.19 | 20,422 |
| Southern Great Plain | Bács-Kiskun | -41.28 | -1.57 | 25,158 |
| | Békés | -41.85 | -1.20 | 19,003 |
| | Csongrád | -25.26 | -1.00 | 26,122 |
| Hungary | | -23.55 | | 660,290 |
| Sum | | | -34.48 | |
| $T_0$ | | | 10.94 | |

Note: In order to avoid reading difficulties we multiplied $T_{GTO}$ values by 1000 and use the terms of millibit.

Table 5: High- and medium-high tech manufacturing vs. knowledge-intensive services and the effects on the mutual information in three dimensions

| Regions | Counties | Knowledge-intensive services ΔT, millibits | Effect on the entropy % | Number of firms among knowledge-intensive service | High- and medium tech ΔT, millibits | Effect on the entropy % | Number of high-tech and medium-tech firms |
|---|---|---|---|---|---|---|---|
| Central Hungary | Budapest | -2.64 | -18.9 | 64,791 | -1.30 | 366.6 | 5,840 |
| | Pest | -0.16 | -51.7 | 8,495 | -1.26 | 1179.0 | 2,551 |
| Western Transdanubia | Gyor-M.-S. | -0.03 | -47.9 | 24,684 | -0.46 | 980.4 | 850 |
| | Vas | -0.01 | -57.0 | 18,563 | -0.36 | 1064.9 | 321 |
| | Zala | -0.01 | -35.2 | 641,143 | -0.15 | 636.0 | 464 |
| Central Transdanubia | Komárom-E. | -0.01 | -58.4 | 29,327 | -0.53 | 1257.2 | 741 |
| | Fejér | -0.02 | -55.8 | 19,888 | -0.54 | 1172.0 | 776 |
| | Veszprém | -0.02 | -46.3 | 25,299 | -0.41 | 957.4 | 645 |
| Southern Transdanubia | Baranya | -0.04 | -16.9 | 27,327 | -0.14 | 329.3 | 624 |
| | Somogy | -0.02 | -21.1 | 25,928 | -0.11 | 298.5 | 394 |
| | Tolna | -0.01 | -29.6 | 24,313 | -0.13 | 632.8 | 349 |
| Northern Hungary | Borsod-A.-Z. | -0.07 | -31.2 | 17,019 | -0.51 | 633.8 | 847 |
| | Heves | -0.01 | -45.3 | 11,995 | -0.29 | 948.5 | 498 |
| | Nógrád | 0.00 | -49.6 | 14,597 | -0.21 | 969.3 | 227 |
| Northern Great Plain | Hajdú-Bihar | -0.03 | -37.8 | 15,286 | -0.30 | 712.4 | 696 |
| | Jász-N.-Sz. | -0.01 | -52.1 | 19,793 | -0.37 | 1102.3 | 557 |
| | Szabolcs-Sz.-B. | -0.03 | -27.8 | 15,956 | -0.23 | 551.7 | 629 |
| Southern Great Plain | Bács-Kiskun | -0.03 | -42.0 | 14,169 | -0.45 | 886.5 | 845 |
| | Békés | -0.03 | -19.7 | 16,074 | -0.16 | 351.0 | 440 |
| | Csongrád | -0.03 | -11.6 | 23,299 | -0.09 | 206.4 | 823 |
| Hungary | | -19.28 | -15.7 | 223,325 | -3.08 | 351.7 | 19,147 |

Table 6: The contribution of high-tech services to knowledge function synergy

| Regions | Counties | KIS mutual information ΔT, mbit | Effect on the entropy % | Number of records | HT-KIS mutual information ΔT, mbit | Effect on the entropy % | Number of records |
|---|---|---|---|---|---|---|---|
| Central Hungary | Budapest | -2.64 | -18.9 | 64,791 | -13.05 | 35.5 | 18,491 |
|  | Pest | -0.16 | -51.7 | 8,495 | -2.75 | -18.7 | 5,019 |
| Western Transdanubia | Gyor-Moson-Sopron | -0.03 | -47.9 | 24,684 | -0.25 | -83.1 | 1,195 |
|  | Vas | -0.01 | -57.0 | 18,563 | -0.39 | -63.3 | 640 |
|  | Zala | -0.01 | -35.2 | 641,143 | -0.24 | -65.2 | 586 |
| Central Transdanubia | Komárom-Esztergom | -0.01 | -58.4 | 29,327 | -0.31 | -76.5 | 794 |
|  | Fejér | -0.02 | -55.8 | 19,888 | -0.67 | -54.1 | 1,211 |
|  | Veszprém | -0.02 | -46.3 | 25,299 | -0.35 | -74.4 | 836 |
| Southern Transdanubia | Baranya | -0.04 | -16.9 | 27,327 | -0.14 | -88.1 | 1,325 |
|  | Somogy | -0.02 | -21.1 | 25,928 | -0.58 | -41.9 | 638 |
|  | Tolna | -0.01 | -29.6 | 24,313 | -0.32 | -49.7 | 517 |
| Northern Hungary | Borsod-Abaúj-Zemplén | -0.07 | -31.2 | 17,019 | -0.81 | -66.0 | 1,387 |
|  | Heves | -0.01 | -45.3 | 11,995 | -0.21 | -78.1 | 668 |
|  | Nógrád | 0.00 | -49.6 | 14,597 | -0.19 | -71.3 | 332 |
| Northern Great Plain | Hajdú-Bihar | -0.03 | -37.8 | 15,286 | -0.38 | -70.4 | 1,225 |
|  | Jász-Nagykun-Szolnok | -0.01 | -52.1 | 19,793 | -0.38 | -64.1 | 709 |
|  | Szabolcs-Szatmár-Bereg | -0.03 | -27.8 | 15,956 | -0.49 | -58.9 | 811 |
| Southern Great Plain | Bács-Kiskun | -0.03 | -42.0 | 14,169 | -0.91 | -42.0 | 1,075 |
|  | Békés | -0.03 | -19.7 | 16,074 | -0.39 | -67.8 | 571 |
|  | Csongrád | -0.03 | -11.6 | 23,299 | -1.82 | 82.0 | 1,383 |
| Hungary |  | -19.28 | -15.7 | 223,325 | -12.02 | -49.0 | 39,415 |